\documentclass[aps,prb,preprint,showpacs,superscriptaddress]{revtex4}

\usepackage{graphicx} %Include figure files

\begin{document}
\draft
\thispagestyle{plain}

\preprint{submitted to Phys. Rev. B.}

\title {Simultaneous valence shift of Pr and Tb ions in (Pr$_{1-y}$Tb$_{y})_{0.7}$Ca$_{0.3}$CoO$_{3}$ around M-I transition}

\author{H. Fujishiro}
\author{T. Naito}
\author{D. Takeda}
\author{N. Yoshida}
\author{T. Watanabe}
\affiliation{Faculty of Engineering, Iwate University, 4-3-5 Ueda, Morioka 020-8551, Japan}

\author{K. Nitta}
\affiliation{Japan Synchrotron Radiation Research Institute, Sayo, Hyogo 679-5198, Japan}

\author{J. Hejtm\'{a}nek}
\author{K. Kn\'{i}\v{z}ek}
\author{Z. Jir\'{a}k}
\affiliation{Institute of Physics, ASCR, Cukrovarnick\'{a} 10, 162 00 Prague 6, Czech Republic}

\date{\today}

\begin{abstract}
   Temperature dependence of the X-ray absorption near-edge structure (XANES) spectra at the Pr $L_{3}$- and Tb $L_{3}$-edges was measured for the (Pr$_{1-y}$Tb$_{y})_{0.7}$Ca$_{0.3}$CoO$_{3}$ system, in which a metal-insulator (MI) and spin-state (SS) transition took place simultaneously at a critical temperature $T_{\rm MI}$. 
   A small increase in the valence of the terbium ion was found below $T_{\rm MI}$, besides the enhancement of the praseodymium valence; the trivalent states, which are stable at room temperature, change to a 3+/4+ ionic mixture at low temperatures. 
   In particular for the $y$=0.2 sample, the average valence determined at 8 K amounts to 3.25+ and 3.03+ for the Pr and Tb ion, respectively. 
   In analogous (Pr$_{1-y}$RE$_{y})_{0.7}$Ca$_{0.3}$CoO$_{3}$ samples (RE=Sm and Eu), in which the MI-SS transition also took place, no valence shift of the RE ion was detected in the XANES spectra at the RE ion $L_{3}$-edge. 
   The role of the substituted RE ion for the Pr-site on the MI-SS transition is discussed. 
\end{abstract}

%74.25.Fy Transport properties
%74.60.Ec Mixed state, critical fields, and surface sheath
%74.25.Qt Flux pinning, flux creep, and flux-line lattice dynamics
%74.72.Bk Y-based cuprates
%74.72.Hs Bi-based cuprates  


\maketitle

\section{Introduction}
   The perovskite cobaltites RECoO$_{3}$ (RE=rare-earth element and Y) show a spin-state (SS) transition of Co$^{3+}$ ions from a low spin state (LS; $t_{2g}^{6}e_{g}^{0}$, $S$=0) to a high spin state (HS; $t_{2g}^{4}e_{g}^{2}$, $S$=2) with increasing temperature, followed by the formation of the metallic state of the intermediate spin state (IS; $t_{2g}^{5}\sigma^{\ast}$, $S$=1) at higher temperatures.~\cite{Jirak2008} 
    The temperature induced spin-state transition and its gradual course indicate a small energy difference $\delta E$ between the crystal-field splitting and Hund coupling energy.~\cite{Korotin1996,Yan2004} 
    The hole-doped systems like La$_{1-x}$Sr$_{x}$CoO$_{3}$ generally show a temperature stable phase with itinerant cobalt states (presumably the mixed IS Co$^{3+}$/ LS Co$^{4+}$ or HS Co$^{3+}$/ IS Co$^{4+}$ configurations) and undergo a ferromagnetic ordering at low temperatures.    
    Most interestingly, some Pr-based cobaltites exhibit a pronounced first-order transition to a low-temperature phase of weakly paramagnetic character and reduced conduction. This transition is referred in the literature as the metal-insulator (MI) transition, because of the large resistivity change making one or two orders of magnitude and its sharpness.
   It was revealed for the first time on Pr$_{0.5}$Ca$_{0.5}$CoO$_{3}$ at $T_{\rm MI}$$\sim90$~K, and was documented, in addition to the step-like resistivity jump, by concomitant anomalies in the magnetic susceptibility, heat capacity and lattice dilatation.~\cite{Tsubouchi2002,Tsubouchi2004} 
    The change in the electronic structure was confirmed at $T_{\rm MI}$ by the photoemission spectroscopy.~\cite{Saito2005} 
    The mechanism of the transition was tentatively ascribed to a spin-state crossover from the itinerant cobalt states to an ordered mixture of localized LS Co$^{3+}$ and LS Co$^{4+}$ ($t_{2g}^{5}e_{g}^{0}$, $S$=1/2) states. 
    Thereafter, the existence of Co$^{3+}$/Co$^{4+}$ ordering was questioned because similar transition was evidenced also in the less doped Pr$_{1-x}$Ca$_{x}$CoO$_{3}$ ($x$=0.3) under high pressures,~\cite{Fujita2004} or in the (Pr$_{1-y}$RE$_{y})_{1-x}$Ca$_{x}$CoO$_{3}$ system (0.2$\le x\le$0.5) with a partial substitution of Pr by smaller RE cations such as Sm, Eu and Y under ambient pressure.~\cite{Fujita2004,Fujita2005,Naito2010} 
    This MI transition accompanied by SS transition appeared to be conditioned not only by the presence of both Pr and Ca ions, but also by a larger structural distortion of the CoO$_{6}$ network, depending on the average ionic radius and size mismatch of perovskite {\it A}-site ions.~\cite{Naito2010} 
    Furthermore, the critical temperature $T_{\rm MI}$ was found to be depressed by the applied magnetic field.~\cite{Marysko2011,Naito}

    An alternative scenario explaining the nature of such specific transition was proposed on the basis of electronic structure calculations exploiting the temperature dependence of the structural experimental data for Pr$_{0.5}$Ca$_{0.5}$CoO$_{3}$.~\cite{Knizek2010a} 
    It appeared that the formal cobalt valence should change below $T_{\rm MI}$ from mixed-valence Co$^{3.5+}$ towards pure Co$^{3+}$ with strong preference for the LS state and, concomitantly, the praseodymium valence should increase simultaneously from Pr$^{3+}$ towards Pr$^{4+}$. 
    The SS transition and electronic localization was thus intuitively interpreted as an analogy of the compositional transition from the ferromagnetic metal of La$_{0.5}$Ca$_{0.5}$CoO$_{3}$ to the diamagnetic insulator of LaCoO$_{3}$.~\cite{Wu2003,Knizek2010b}
    Shortly afterwards, the theoretical hypothesis about the crucial role of variable praseodymium valence was experimentally supported by observation of a Schottky peak in the low-temperature specific heat of (Pr$_{1-y}$Y$_{y})_{0.7}$Ca$_{0.3}$CoO$_{3}$ ($y$=0.075 and 0.15),~\cite{Hejtmanek2010} which proved a stabilization of Kramers Pr$^{4+}$ ions in the low-temperature phase.  
    
    Most recently, we directly confirmed the praseodymium valence shift, realized in fact as a mixture of the Pr$^{3+}$ and Pr$^{4+}$ states, from the X-ray absorption near-edge structure (XANES) spectra at the Pr $L_{3}$-edge for the same (Pr$_{1-y}$Y$_{y})_{0.7}$Ca$_{0.3}$CoO$_{3}$ samples.~\cite{Fujishiro2012}
    It has been found that the average valence of the praseodymium ion increases below room temperature from the common value 3.0+, undergoes a steepest change at $T_{\rm MI}$ and reaches finally 3.15+ and 3.27+ at 8 K for the $y$=0.075 and 0.15 samples, respectively. 
    Interestingly, these values are consistent with those estimated by quantitative analysis of the entropy associated with the Schottky peak dominating the low-temperature specific heat. 
    For Pr$_{0.5}$Ca$_{0.5}$CoO$_{3}$, similar valence shifts from 3.0+ at 300 K to 3.15+ at 10 K were obtained by other authors using XANES at Pr $L_{3}$ and $M_{4,5}$ edges.~\cite{Garcia-Munoz2011} 
    In addition, a change of cobalt spin states has been evidenced by the analysis of two complementary synchrotron X-ray spectroscopic techniques.~\cite{Herrero2012}
    
   In the present work, we apply XANES spectroscopy measurements to the analysis of (Pr$_{1-y}$RE$_{y})_{0.7}$Ca$_{0.3}$CoO$_{3}$ systems (RE=Tb, Sm and Eu), in which the MI-SS transition between 50--140 K was obtained.~\cite{Fujita2005,Naito2010} 
   The choice of substituting RE ions was motivated by the fact that terbium ion is known to adopt the mixed valence state Tb$^{3+}$/Tb$^{4+}$ in several compounds.~\cite{Largeau1998} 
   The Sm and Eu ions are known to exist also in various valence states. 
   As an example, in the Sm-based filled skutterudite, SmOs$_{4}$Sb$_{12}$, Sm ions possess a mixed valence state between Sm$^{2+}$ and Sm$^{3+}$, which can be detected by XANES.~\cite{Mizumaki2007} 
   In the Eu-contained niobates such as Eu$_{2}$Nb$_{5}$O$_{9}$ and EuNbO$_{3}$, a mixed valence state between Eu$^{2+}$ and Eu$^{3+}$ was detected by photoelectron spectroscopy.~\cite{Felser1998} 
   In the present study, we thus show a detailed temperature dependence of the XANES spectra around the RE $L_{3}$-edges measured using the bulk sensitive transmission method, and derive the RE ion states from the spectra.

\section{Experimental}
      Polycrystalline samples (Pr$_{1-y}$Tb$_{y})_{0.7}$Ca$_{0.3}$CoO$_{3}$ ($y$=0, 0.1, 0.2), (Pr$_{1-y}$Sm$_{y})_{0.7}$Ca$_{0.3}$CoO$_{3}$ ($y$=0.2, 0.3, 0.4), and (Pr$_{1-y}$Eu$_{y})_{0.7}$Ca$_{0.3}$CoO$_{3}$ ($y$=0.15, 0.2, 0.3) were fabricated by a solid-state reaction.
      Raw powders of Pr$_{6}$O$_{11}$, Tb$_{4}$O$_{7}$, Sm$_{2}$O$_{3}$, Eu$_{2}$O$_{3}$, Co$_{3}$O$_{4}$, and CaCO$_{3}$ were weighted with proper molar ratios and ground using an agate mortar and pestle for 1 h. 
     Mixed powders were calcined at 1000 $^\circ$C for 24 h in air. 
     Then they were pulverized, ground, and pressed into pellets of 20 mm diameter and 4 mm thickness. Pellets were sintered at 1200 $^\circ$C for 24 h in 0.1 MPa flowing oxygen gas. 
     The measured relative densities of each sample were greater than 90$\%$. 
     Powder X-ray diffraction patterns were taken for each sample using Cu {\it K}{\it $\alpha$} radiation; the samples were confirmed to have a single-phase orthoperovskite {\it Pbnm} structure. 
     The oxygen stoichiometry was checked on the $y$=0 sample (Pr$_{0.7}$Ca$_{0.3}$CoO$_{3}$) using the high resolution neutron diffraction, and practically ideal oxygen content $2.99\pm0.01$ has been found.
     For the XANES measurements, a small amount of the samples were pulverized, mixed with boron nitride powder (99.9$\%$) with proper molar ratios in order to optimize absorption, and pelletized 6 mm in diameter and 0.5 mm in thickness.

    Each XANES spectrum of the samples was measured at BL01B1 of SPring-8 in Japan. 
    The beam was monochromatized using a Si(111) double-crystal monochromator. 
    The spectra were recorded in the transmission mode with the detectors of the ionization chambers and obtained at various temperatures from 8 to 300 K using a cryocooler. 
    The measurements were performed upon the heating run. The energy resolution was within 1.5 eV around $E$=8 keV.

  To determine the mixed Pr$^{3+}$/Pr$^{4+}$ contents at low temperatures for (Pr$_{1-y}$Tb$_{y})_{0.7}$Ca$_{0.3}$CoO$_{3}$, the calibration line was obtained using XANES spectra of Pr$_{6}$O$_{11}$ as Pr$^{3.667+}$ and (Pr$_{1-y}$Tb$_{y})_{0.7}$Ca$_{0.3}$CoO$_{3}$ as Pr$^{3.0+}$ measured at 300 K, which were performed similarly to the previous paper.~\cite{Fujishiro2012} 
  The recorded XANES spectra were modeled by the sum of three Lorentzian functions and one arctangent function representing the step like edge of continuum excitations. 
  One Lorentzian function (peak A: 5966 eV) shows an excitation from 2$p_{3/2}$ to 4$f^{2}5d^\ast$, which represents Pr$^{3.0+}$ ions. 
  The other Lorentzian functions (peak B2: 5969 eV and peak B1: 5979 eV) show the excitations from 2$p_{3/2}$ to 4$f^{2}\underline{L}5d^\ast$ and to 4$f^{1}5d^\ast$, $\underline{L}$ being a ligand hole in the O 2$p$ orbital, both of which represent Pr$^{4+}$ ions.~\cite{Yamaoka2008,Bianconi1987} 
  The energy differences between peaks A, B2, and B1 were fixed according to results of Hu {\it et al.}~\cite{Hu1994} 
  The curve fittings were performed in the energy range from 5944 to 5985 eV using Athena software.~\cite{Ravel2005} 

   To estimate the mixed Tb$^{3+}$/Tb$^{4+}$ contents, a similar procedure was applied. 
   A comparative measurement of Tb$_{4}$O$_{7}$ and (Pr$_{1-y}$Tb$_{y})_{0.7}$Ca$_{0.3}$CoO$_{3}$ was carried out at 300K as standards of Tb$^{3.5+}$ and Tb$^{3.0+}$, respectively. 
   The recorded XANES spectra for Tb $L_{3}$ line were modeled by the sum of two Lorentzian functions (peak C: 7518 eV for Tb$^{3+}$ (4$f$$^{8}$) and peak D: 7528 eV for Tb$^{4+}$ (4$f$$^{7}$)) and one arctangent function. 
   For simplicity, the peak related to the ligand hole state 4$f$$^{8}\underline{L}$ was omitted because of its weaker intensity as reported by Dexpert $et   al.$~\cite{Dexpert1987} 
   For (Pr$_{0.7}$Sm$_{0.3})_{0.7}$Ca$_{0.3}$CoO$_{3}$, and (Pr$_{0.85}$Eu$_{0.15})_{0.7}$Ca$_{0.3}$CoO$_{3}$ samples, the XANES spectra were measured around Pr $L_{3}$ (5966 eV), Sm $L_{3}$ (6720 eV) and Eu $L_{3}$ (6981 eV) edges, respectively.

\section{Results and Discussion}
\subsection{Electrical and magnetic properties in (Pr$_{1-y}$Tb$_{y})_{0.7}$Ca$_{0.3}$CoO$_{3}$}

    The electrical resistivity {\it $\rho$}($T$) and magnetic susceptibility {\it $\chi$}($T$) of the (Pr$_{1-y}$Tb$_{y})_{0.7}$Ca$_{0.3}$CoO$_{3}$ ($y$=0, 0.1, 0.2) samples are presented in Figs. 1(a) and 1(b). 
    The resistivity measured on the $y$=0 sample is only weakly temperature dependent, without any sign of phase transition. 
    There is a shallow minimum in the high temperature range (see the inset of Fig. 1(a)), then the resistivity increases toward low temperatures but still extrapolates to finite values of about $10$~mOhm.cm at zero K. 
    This is indicative for metallic nature of the sample ground state, in the sense of Moebius' criterion for inhomogeneous or granular systems, $d(ln \rho)/d(ln T) \rightarrow 0$. 
    Another signature for intrinsic metallicity of our ceramic sample Pr$_{0.7}$Ca$_{0.3}$CoO$_{3}$ is the apparent activation energy, $E_A=k.d(ln\rho)/d(1/T)$, which does not exceed the thermal energy $k_BT$ in the entire temperature range. 
      On the other hand, the Tb-substituted samples show features typical for MI transition similarly to Pr$_{0.5}$Ca$_{0.5}$CoO$_{3}$.
      While the behavior in the high-temperature range is analogous to that of $y$=0, the low-temperature resistivity measured on virgin sample exhibits a sharp jump below the characteristic temperatures of $T_{\rm MI}$$\sim75$ and 145~K for $y$=0.1 and 0.2, respectively. 
      The tendency to resistivity saturation at the lowest temperatures shows that no real gap at Fermi level is opened in the low-temperature phase; the resistivity jump at $T_{\rm MI}$ is thus due to setting of E$_F$ close to the mobility edge, which has been confirmed by the regime of variable range hopping found for analogous system (Pr$_{1-y}$Y$_{y})_{0.7}$Ca$_{0.3}$CoO$_{3}$\cite{Hejtmanek2010}). 
    On the heating run, {\it $\rho$}($T$) shows a similar transition, but the resistivity value above $T_{\rm MI}$ is not restored and is systematically larger. 
    Further temperature cycling is then reproducible with minimal hysteresis.
    The irreversibility of virgin state can be related to micro cracks in the samples due to the large volume contraction when passing through the transition temperature.~\cite{Naito2010} 
    Let us note that our unpublished experiments on the prototypical compound Pr$_{0.5}$Ca$_{0.5}$CoO$_{3}$ showed similar irreversibility not only in the electrical resistivity but also in thermal conductivity and diffusivity, proving directly the reduction of the heat transport due to damaged microstructure. The next cycling led to reversible behavior with small thermal hysteresis of about 1~K, in close agreement with the thermal hysteresis detected in magnetic susceptibility data.  
    It can be added that the hysteretic behavior of resistivity may change depending on the measured samples; Tsubouchi reported for Pr$_{0.5}$Ca$_{0.5}$CoO$_{3}$ a large irreversibility between the virgin cooling and subsequent heating,~\cite{Tsubouchi2002} which is the same case as ours, but Herrero-Mart\'{i}n reported less hysteretic resistivity.~\cite{Herrero2012}

    The transitions in the (Pr$_{1-y}$Tb$_{y})_{0.7}$Ca$_{0.3}$CoO$_{3}$ system are further evidenced in the plot of magnetic susceptibility.  
    For the Tb free sample $y$=0 with absence of MI transition, {\it $\chi$}($T$) steeply increases to large values at low temperatures in accordance with the ferromagnetic ground state of this compound ($T_{\rm C}$$\sim55$~K). 
    The Tb substituted samples exhibit a marked drop of magnetic susceptibility at nearly the same temperature as $T_{\rm MI}$ decuced from {\it $\rho$}($T$), which is a strong sign that cobalt ions transform to LS states concomitantly with MI transition.~\cite{Hejtmanek2010} 
    These characteristics of the Tb-substituted samples are the same as those observed for (Pr$_{1-y}$Y$_{y})_{0.7}$Ca$_{0.3}$CoO$_{3}$ and other (Pr$_{1-y}$RE$_{y})_{0.7}$Ca$_{0.3}$CoO$_{3}$ systems.~\cite{Fujita2005,Naito2010} 
    It is worth mentioning that, in distinction to Y and Sm substituted samples, the low-temperature susceptibility increases with $y$, which should be related to large moment of Tb ion independently on its trivalent or tetravalent state. 
    (Tb$^{3+}$ with 4$f$$^{8}$ configuration is a specific case of non-Kramers ion. 
    Although its eigenstates are nonmagnetic in the low-symmetry crystal field of {\it {Pbnm}} perovskite structure, the ground and first excited singlets are nearly degenerate, forming a magnetic quasi-doublet.~\cite{Gruber2008})

\subsection{XANES measurements of Pr $L_{3}$- and Tb $L_{3}$-edges}
     Figure 2(a) shows the temperature dependence of the XANES spectra at the Pr $L3$-edge for the (Pr$_{0.8}$Tb$_{0.2})_{0.7}$Ca$_{0.3}$CoO$_{3}$ sample. 
     Two main peaks situated at 5966 and 5979 eV (named peaks A and B1) originate from the Pr $2p \rightarrow 5d$ transitions. 
     At 300 K, the Pr$^{3+}$ ($4f^{2}$) sites essentially contribute to the peak A, with a little component at peak B2, which is caused presumably by multiple scattering, commonly treated in the theoretical simulations of XANES.~\cite{Fujishiro2012} 
     At temperatures close to $T_{\rm MI}$$\sim145$~K, the shape of the XANES spectra changes markedly; the intensity of peak B1 increases notably and shifts to lower energy side, while the intensity of peak A decreases.  
     At the same time, a new component (peak B2 at 5969 eV) can be resolved on its high-energy slope. Since the B1 and B2 peaks are manifestations of Pr$^{4+}$ states, originating in particular of the configurations $4f^{1}$ and $4f^{2}\underline{L}$,~\cite{Yamaoka2008,Bianconi1987,Hu1994}  the observed changes confirm that the average valence of the Pr ions in the (Pr$_{1-y}$Tb$_{y})_{0.7}$Ca$_{0.3}$CoO$_{3}$ increases from 3+ toward 4+ below $T_{\rm MI}$, consistently with those reported for (Pr$_{1-y}$Y$_{y})_{0.7}$Ca$_{0.3}$CoO$_{3}$ by us~\cite{Fujishiro2012} or for Pr$_{0.5}$Ca$_{0.5}$CoO$_{3}$ by Garc\'{i}a-Mu\~{n}oz {\it {et al.}}~\cite{Garcia-Munoz2011} 
      In Fig. 2(a), the first oscillation around 6000 eV can be seen in the XANES spectra, where the broad peak shifts to high-energy side with decreasing temperature. 
      These results qualitatively suggest that the Pr-O distance shortens with decreasing temperature.
      The detailed analysis must be performed using extended X-ray absorption fine structure (EXAFS).

      Figure 2(b) shows the temperature dependence of the XANES spectra at the Tb $L_{3}$-edge for the same (Pr$_{0.8}$Tb$_{0.2})_{0.7}$Ca$_{0.3}$CoO$_{3}$ sample. 
      Two main peaks situated at 7518 and 7529 eV (named peaks C and D) originate from the Tb $2p \rightarrow 5d$ transitions. 
      The spectrum at 300 K is characteristic for the Tb$^{3+}$ ($4f^{8}$) valence, contributing to the main peak C. 
     The form of XANES spectra is changing around $T_{\rm MI}$$\sim145$~K. 
     There is a small drop of the intensity of peak C, which is a fingerprint of Tb$^{3+}$, and at the same time, peak D increases slightly and shifts to lower energy.
       The peak D is a manifestation of Tb$^{4+}$ states, namely, of the configuration $4f^{7}$.~\cite{Dexpert1987,Kalkowski1988} 
      The observed behavior thus suggests that the valence of the Tb ions increases below $T_{\rm MI}$, simultaneously with the valence shift of Pr ions.
      In Fig. 2(b), a slight peak shift of the first oscillation around 7552 eV can be also observed with decreasing temperature. 

\subsection{Valence shifts of the Pr and Tb ions}
         In order to determine the valence shift of Pr ion quantitatively, the XANES spectra were fitted to a sum of three Lorentzian functions (peak A for Pr$^{3+}$ and peaks B1 and B2 for Pr$^{4+}$) and one arctangent function, as indicated in ref. 16. 
         Then the valence of Pr ions in the Tb substituted samples was deduced from the intensity ratio {\it I}$_{B1}$/{\it I}$_{A}$ of the B1 spectral peak to the A spectral peak using the calibration line that took into account the actual {\it I}$_{B1}$/{\it I}$_{A}$ values for Pr$^{3.0+}$ and that for Pr$^{3.667+}$ (not shown). 
         Figure 3(a) shows the temperature dependence of the XANES spectra at the Pr $L_{3}$-edge for pure Pr$_{0.7}$Ca$_{0.3}$CoO$_{3}$ with ferromagnetic metallic ground state (see Fig. 1). 
         Apart of the main peak A, there is again a small bump at around 5980 eV, originating in the multiple scattering as mentioned above. 
         The same feature has been observed for another metallic compound Pr$_{0.55}$Ca$_{0.45}$CoO$_{3}$,~\cite{Garcia-Munoz2011} and seems thus to be a general manifestation of trivalent praseodymium in the mixed Co$^{3+}$/Co$^{4+}$ systems. 
         Importantly, no spectral change of the Pr $L_{3}$-edge is observed down to 8 K, which suggests that the Pr$^{3+}$ valence in Pr$_{0.7}$Ca$_{0.3}$CoO$_{3}$ remains temperature independent.
         
           Figure 3(b) shows the XANES spectrum for Tb$_{4}$O$_{7}$ at 300 K, which was fitted to a sum of two Lorentzian functions (peak C for Tb$^{3+}$ and peak D for Tb$^{4+}$) and one arctangent function. 
           The measurement was also performed down to low temperatures and since no relevant spectroscopic changes were detected, the spectrum, in particular the intensity ratio {\it I}$_{D}$/{\it I}$_{C}$, was further considered as a standard for Tb$^{3.5+}$.  
           The valence of Tb ions in the (Pr$_{1-y}$Tb$_{y})_{0.7}$Ca$_{0.3}$CoO$_{3}$ samples was simply deduced from observed intensity ratio {\it I}$_{D}$/{\it I}$_{C}$ of the D to C spectral peaks,~\cite{Yamaoka2008} taking the data at 300 K as a standard of Tb$^{3.0+}$. 
           This has been done with use of the calibration line in the inset of Fig. 3(b); {\it I}$_{D}$/{\it I}$_{C}$=0.856 for Tb$^{3.5+}$ in Tb$_{4}$O$_{7}$ and {\it I}$_{D}$/{\it I}$_{C}$=0.136 for Tb$^{3.0+}$ in (Pr$_{0.8}$Tb$_{0.2})_{0.7}$Ca$_{0.3}$CoO$_{3}$.

       The fit of XANES spectra is exemplified in Fig. 4 using the data for the (Pr$_{0.8}$Tb$_{0.2})_{0.7}$Ca$_{0.3}$CoO$_{3}$ sample at 8 K, where the valence change achieves a maximum. 
       It turns out that the XANES spectrum at Pr $L_{3}$-edge is well reproduced within the energy range from 5944 to 5985 eV including the peaks related to Pr$^{3+}$ and Pr$^{4+}$. 
       The B2 component at the slope of main peak A is also resolved and the presence of Pr$^{3+}$/Pr$^{4+}$ mixture is thus obvious. 
       Figure 4(b) shows the fit of the XANES spectrum at Tb $L_{3}$-edge for the same sample, also at 8 K. 
       The XANES spectrum cannot be reproduced without considering both characteristic peaks of Tb$^{3+}$ and Tb$^{4+}$ (features C and D, respectively).
       
          The results based on the curve fitting of XANES spectra are summarized in Figs. 5(a) and 5(b), which show, respectively, the temperature dependence of the valence of Pr and Tb ions in the (Pr$_{1-y}$Tb$_{y})_{0.7}$Ca$_{0.3}$CoO$_{3}$ samples. 
          Contrary to the sharp MI-SS transitions detected by the electric resistivity and magnetic susceptibility, the average Pr valence changes on cooling from 300 K gradually, though with the steepest increase at $T_{\rm MI}$. 
          Finally at the lowest temperatures, the valence  reaches final values of 3.19+ and 3.25+ for the $y$=0.1 and 0.2 samples, respectively. 
          The uncertainty of the estimated valence values is within 1$\%$ of the absolute value, which arises from the arbitrariness of parameters used in the arctangent and Lorentzian functions. 
          Comparing the results for (Pr$_{1-y}$Tb$_{y})_{0.7}$Ca$_{0.3}$CoO$_{3}$ and those for the (Pr$_{1-y}$Y$_{y})_{0.7}$Ca$_{0.3}$CoO$_{3}$ samples ($y$=0.075 and 0.15)~\cite{Fujishiro2012} also presented in Fig. 5(a), one finds that the enhancement of the Pr valence systematically increases with increasing contents of the substitution with smaller Tb or Y ions, {\it {i.e.}} with decreasing of the mean size of the $A$-site cation in the perovskite structure.
   
     As the Tb valence in (Pr$_{1-y}$Tb$_{y})_{0.7}$Ca$_{0.3}$CoO$_{3}$ is concerned, it changes gradually around $T_{\rm MI}$ and reaches final values of 3.01+ and 3.03+ at 8 K for the $y$=0.1 and 0.2 samples, respectively (see Fig. 5(b)). 
     We note, however, that the increase in the Tb valence is about one order of magnitude smaller and its ``gradual'' character is even more pronounced than in the case of Pr valence.

\subsection{XANES measurements at RE edges for (Pr$_{1-y}$RE$_{y})_{0.7}$Ca$_{0.3}$CoO$_{3}$ (RE=Sm, Eu)}
       To illustrate the valence states of the substituted Sm and Eu ions in (Pr$_{1-y}$RE$_{y})_{0.7}$Ca$_{0.3}$CoO$_{3}$, the temperature dependent XANES spectra at the Sm $L_{3}$-edge and Eu $L_{3}$-edge are compiled in Figs.  6(a) and 6(b). 
       The insets present the data on electrical resistivity {\it $\rho$}($T$) on the heating run, which demonstrate the existence of MI transition with characteristic temperatures $T_{\rm MI}$ of 85 K and 55 K for the RE=Sm and Eu samples, respectively. 
       
       For the samarium substituted sample, although the intensity of the main peak of Sm$^{3+}$ at 6720 eV slightly decreased with decreasing $T$, no isomeric shift or new spectral feature is detected. 
       This result suggests that the Sm valence remains essentially as 3.0+ over the entire temperature range. 
       For the europium substituted sample, the spectrum does not change at all down to 8 K and, consequently, the valence of the Eu ion also remains 3.0+.

\subsection{Discussion}
        The unusual MI-SS transition occurring in some Pr-based cobaltites is intimately connected with a charge transfer between praseodymium and cobalt sites. 
        Such process is enabled by energy closeness of the Pr$^{3+}$/Pr$^{4+}$ valence states. 
        In the previous subsections, a possibility of more complex valence equilibria has been probed on (Pr$_{1-y}$RE$_{y})_{0.7}$Ca$_{0.3}$CoO$_{3}$ systems with the RE=Tb, Sm, Eu substitutions at Pr-sites, in which mixed valence states, RE$^{3+}$/RE$^{4+}$ or RE$^{3+}$/RE$^{2+}$, can eventually also occur. 
        Indeed, a small but unquestionable increase in the valence of RE ion is observed for the Tb substituted samples below $T_{\rm MI}$, besides the substantial enhancement of the Pr ion valence to ~3.25+. 
        On the other hand, no temperature variation of the Sm and Eu valence state is detected for such RE substituted samples. 
            This means that hypothetical redox reaction between rare earths, like Pr$^{3+}$+Sm$^{3+}$ $ \rightarrow  $ Pr$^{4+}$+Sm$^{2+}$, does not play any role both above and below the MI transition.   
          
          To demonstrate the effect of particular rare earths, the transition temperature $T_{\rm MI}$ is plotted in Fig. 7(a) as a function of the average ionic radius of the perovskite $A$-site, $\langle r_{A} \rangle$. 
          The experimental results for the RE=Y system ($y$=0.075, 0.1 and 0.2) are also shown.
          It is seen that the interrelation between $T_{\rm MI}$ and $\langle r_{A} \rangle$ is not universal, the curves gradually shift to the large $\langle r_{A} \rangle$ side with decreasing ionic radius of RE. 
          In this comparison, the curve for RE=Tb is not peculiar, which can be understood, since the actual shift of Tb valence is fairly small (to 3.03+ in maximum). 
          Nonetheless, the possibility of Tb$^{3+}$/Tb$^{4+}$ crossover in cobaltites is confirmed and additional studies aiming to exemplify this valence change are under way.  
          As a final note to Fig. 7(a) we may mention that the $T_{\rm MI}$ vs $\langle r_{A} \rangle$ relation is applicable only for the (Pr$_{1-y}$RE$_{y})_{0.7}$Ca$_{0.3}$CoO$_{3}$ systems and does not hold for the intensively studied system Pr$_{0.5}$Ca$_{0.5}$CoO$_{3-\delta}$, in which the Co, Pr valences and $T_{\rm MI}$ strongly depend on the actual oxygen content.~\cite{Tsubouchi2002, Fujita2004, Chichev2007, Tong2009, Garcia-Munoz2011}
                    
             Finally, in Fig. 7(b) we plot the formal valence of Co ions in the (Pr$_{1-y}$RE$_{y})_{0.7}$Ca$_{0.3}$CoO$_{3}$ at 8 K as a function of $\langle r_{A} \rangle$, calculated based on the Pr or Tb valences estimated from XANES data. 
             This is enabled by the fact that, unlike  Pr$_{0.5}$Ca$_{0.5}$CoO$_{3-\delta}$, the Pr$_{0.7}$Ca$_{0.3}$CoO$_{3}$ derived systems are prepared generally as stoichiometric ones, as proved for selected samples by thermogravimetry or neutron diffraction method. 
             The oxygen stoichiometry can be inferred also indirectly, comparing the electric transport properties,  as shown in our previous paper on the Y-substituted samples.~\cite{Hejtmanek2010}
             The results in Fig. 7(b) show that the doping level of the cobalt subsystem decreases from 0.30 hole per f. u. to about 0.15 hole per f. u., similar for Tb content of $y$=0.1 and 0.2. Analogous data for other RE system studied by us are also presented.

\section{Conclusion}
         Temperature dependence of the X-ray absorption near-edge structure (XANES) spectra at the Pr $L_{3}$ and RE $L_{3}$ edges was measured for the (Pr$_{1-y}$RE$_{y})_{0.7}$Ca$_{0.3}$CoO$_{3}$ samples (RE=Tb, Sm and Eu), in which a metal-insulator (MI) and spin-state (SS) transition took place simultaneously at a critical temperature $T_{\rm MI}$. 
         The important experimental results and conclusions are summarized as follows.
         
(1)	In all the studied Pr-based cobaltites, the trivalent praseodymium ions change to a Pr$^{3+}$/Pr$^{4+}$ mixture below $T_{\rm MI}$, which anticipates that, at the same time, the formal cobalt valence should decrease from 3.3+ closer to 3.0+.

(2)	Besides the enhancement of the Pr valence, a small increase in the valence of the Tb ion is found in the (Pr$_{1-y}$Tb$_{y})_{0.7}$Ca$_{0.3}$CoO$_{3}$ samples below $T_{\rm MI}$. 
      In the $y$=0.2 sample, the average valence determined at 8 K makes 3.03+ and 3.25+ for the Tb and Pr ion, respectively. 
      The calculated Co valence is 3.156+ in this case. 
      It should be noted that these results do not necessarily mean very different tendencies of the two rare-earth ions for tetravalent states, since the final valence balance in such chemically inhomogeneous systems depends also on steric factors, including the size mismatch at perovskite $A$-sites.
      
(3)	The observed valence shift of Tb ion indicates the energy closeness of trivalent and tetravalent states in this rare earth. 
      This fact suggests that the unusual MI-SS transition need not be unique for the Pr-based cobaltites but can be expected also for the Tb-based ones.
      
(4)	In the (Pr$_{1-y}$RE$_{y})_{0.7}$Ca$_{0.3}$CoO$_{3}$ samples (RE=Sm and Eu) with similar MI-SS transition, no valence shift of the RE ion is detected at the RE ion absorption edge in the XANES spectra. 
      This suggests that Eu$^{3+}$ and Sm$^{3+}$ are stable in the system over the entire temperature range, and may signify that the energy closeness of trivalent and divalent states in these rare earths has no effect on the praseodymium/cobalt valence equilibria.

\section*{Acknowledgments}
     The synchrotron radiation experiments were performed at the BL01B1 of SPring-8 with approval of Japan Synchrotron Radiation Research Institute (JASRI) (Proposals Nos. 2011A1060, 2011B1075 and 2012A1118). 
     Part of the work was performed under the financial support of a Grant-in-Aid for Scientific Research (No. 24540355) from the Ministry of Education, Culture, Sports, Science and Technology, Japan, and the Grant Agency of the Czech Republic within the Project No. 204/11/0713.

%references

%Figures
\newpage

\begin{figure}[tb]
\includegraphics*[width=0.9\linewidth]{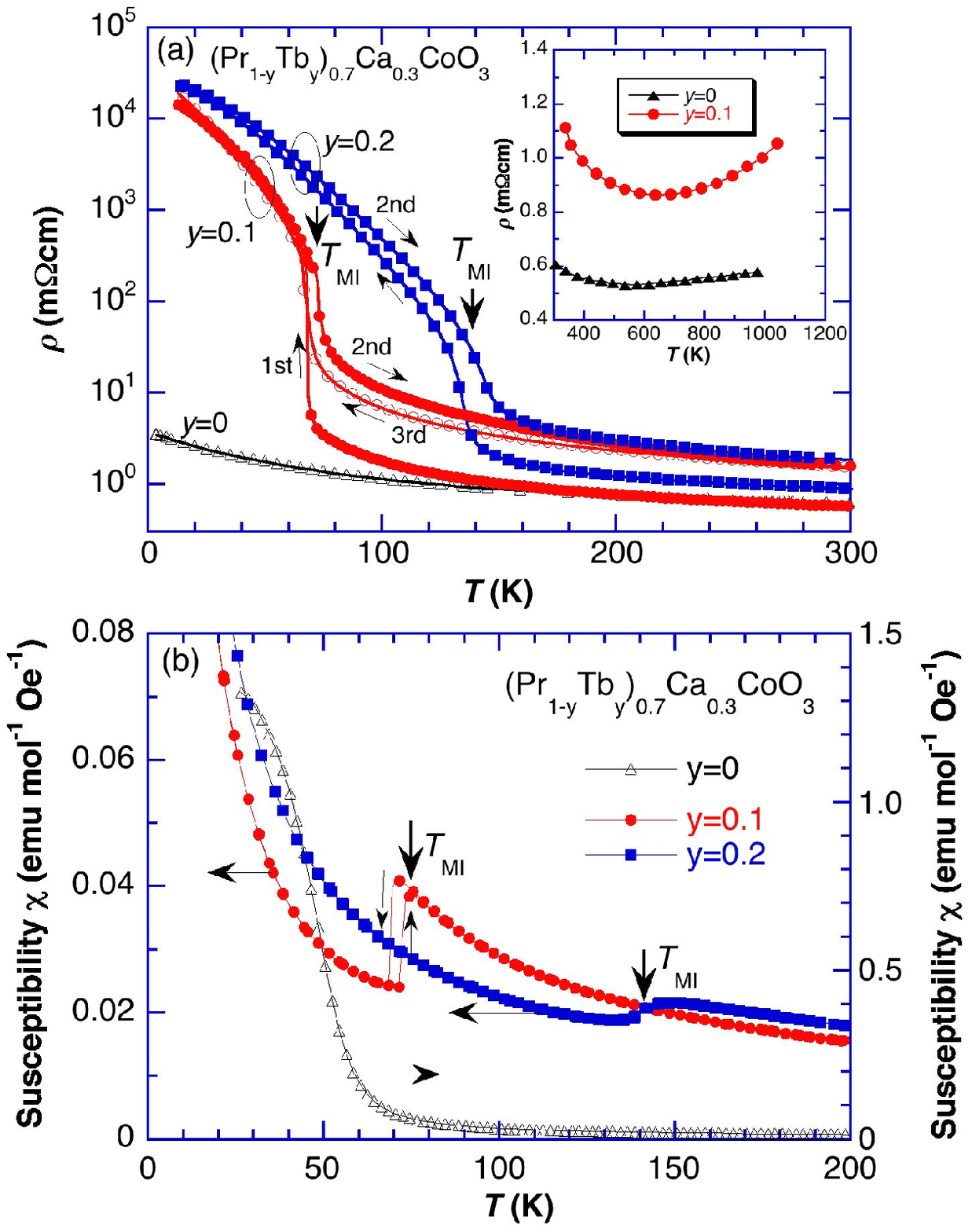}
\caption{(a) Temperature dependence of the electrical resistivity {\it $\rho$}($T$) for the (Pr$_{1-y}$Tb$_{y})_{0.7}$Ca$_{0.3}$CoO$_{3}$ samples ($y$=0, 0.1 and 0.2). High temperature {\it $\rho$}($T$) of the $y$=0 and 0.1 samples were also shown in the inset. For the $y$=0.1 sample, {\it $\rho$}($T$) of the cooling run (1st) for the virgin sample, the subsequent heating run (2nd) and the cooling run (3rd) are shown (see text).
(b) Temperature dependence of the magnetic susceptibility {\it $\chi$}($T$) for the same samples.
}
\label{fig:1}
\end{figure}

\newpage

\begin{figure}[tb]
\includegraphics*[width=0.8\linewidth]{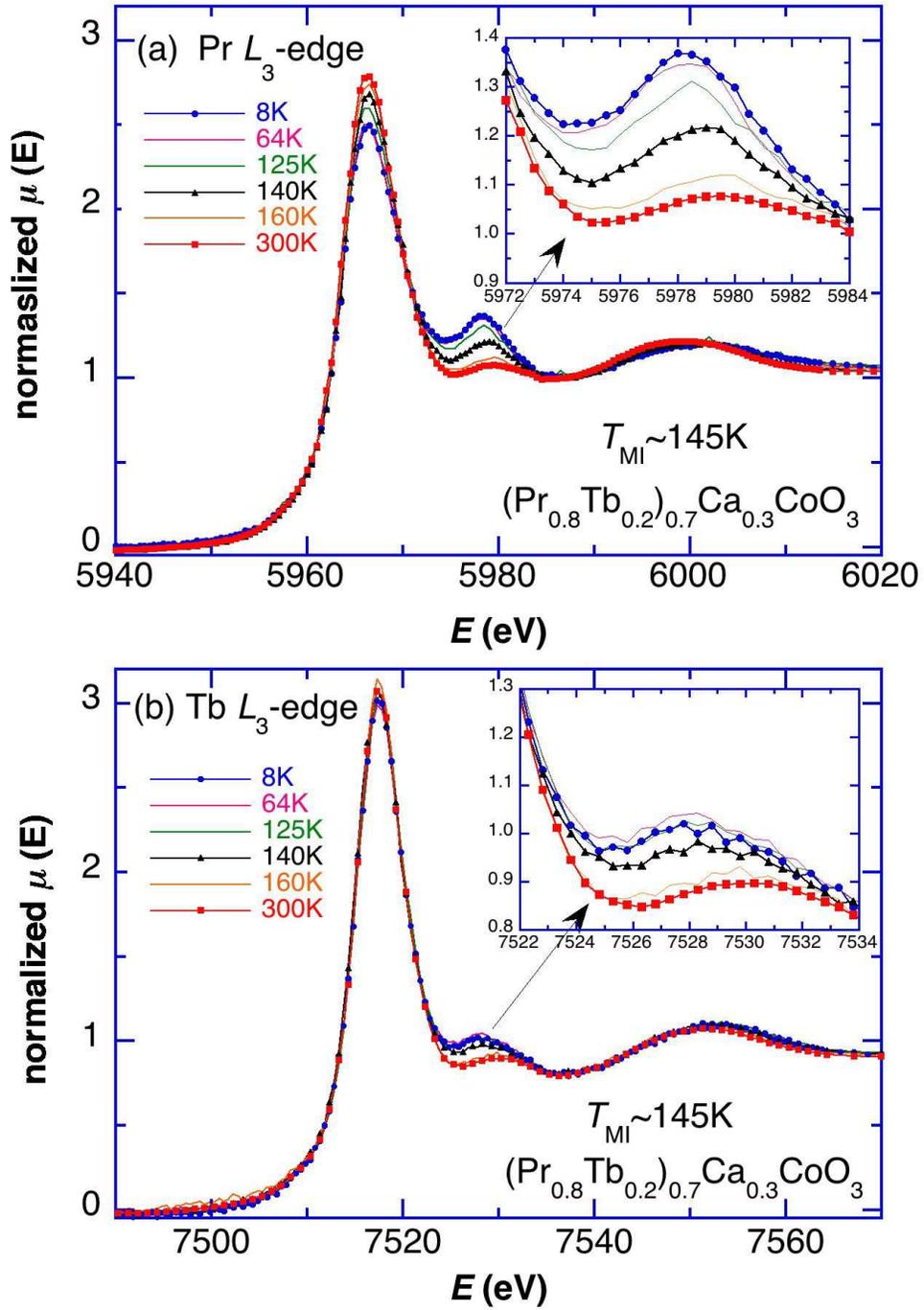}
\caption{Temperature dependence of the XANES spectra at the (a) Pr $L_{3}$-edge and (b) Tb $L_{3}$-edge for the (Pr$_{0.8}$Tb$_{0.2})_{0.7}$Ca$_{0.3}$CoO$_{3}$ sample. Insets show the magnification around the 4$f$$^{1}$ and 4$f$$^{7}$ spectra related with the Pr$^{4+}$ and Tb$^{4+}$ ions, respectively.
}
\label{fig:2}
\end{figure}

\newpage

\begin{figure}[tb]
\includegraphics*[width=0.8\linewidth]{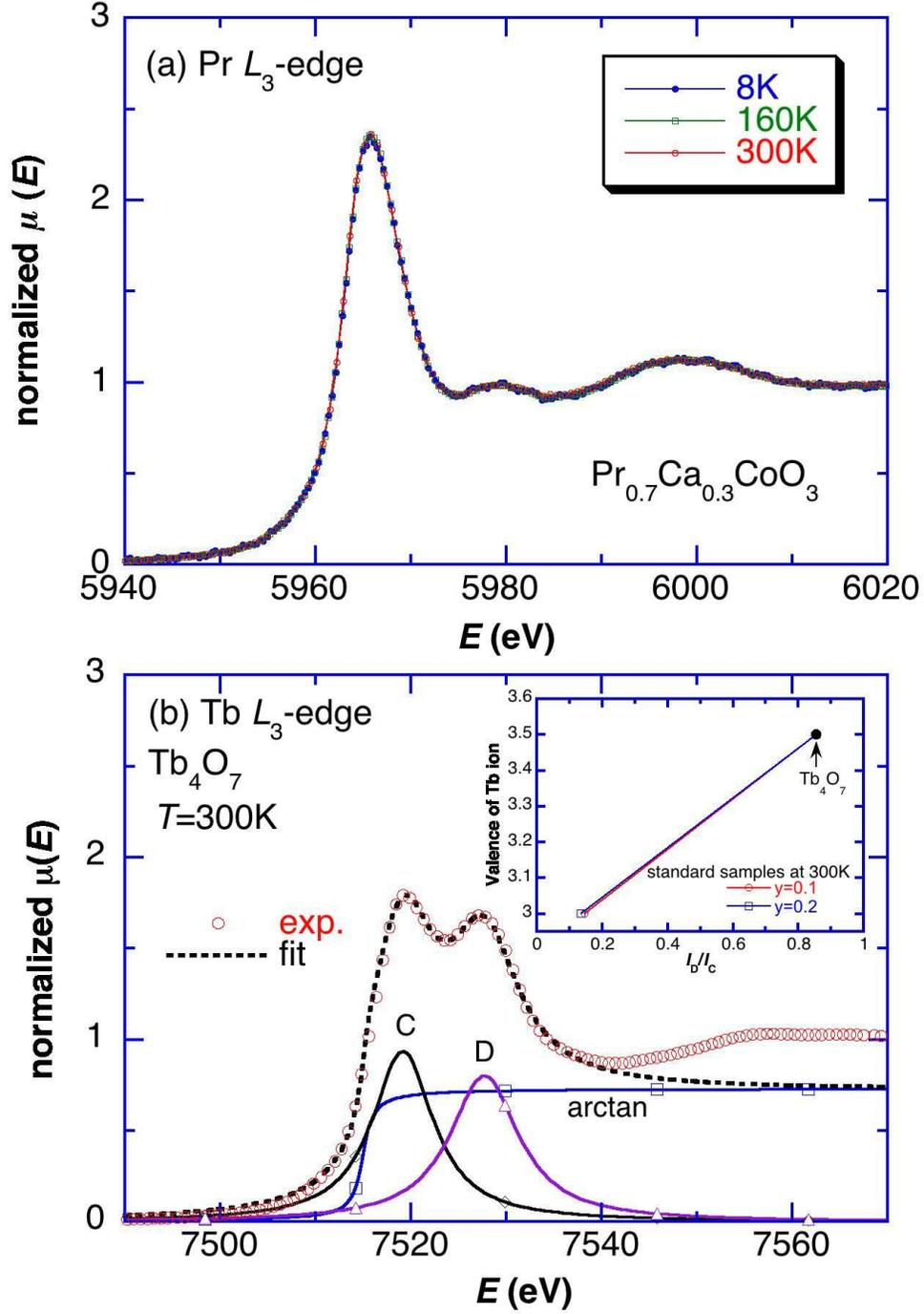}
\caption{(a) Temperature dependence of the XANES spectra at the Pr $L_{3}$-edge for Pr$_{0.7}$Ca$_{0.3}$CoO$_{3}$. (b) The XANES spectrum for Tb$_{4}$O$_{7}$ at 300 K, fitted by a sum of two Lorentzian functions (peak C for Tb$^{3+}$ and peak D for Tb$^{4+}$) and one arctangent function. The inset shows the calibration lines to determine the valence of the Tb ion in the (Pr$_{1-y}$Tb$_{y})_{0.7}$Ca$_{0.3}$CoO$_{3}$ samples ($y$=0.1, 0.2). 
}
\label{fig:3}
\end{figure}

\newpage

\begin{figure}[tb]
\includegraphics*[width=0.8\linewidth]{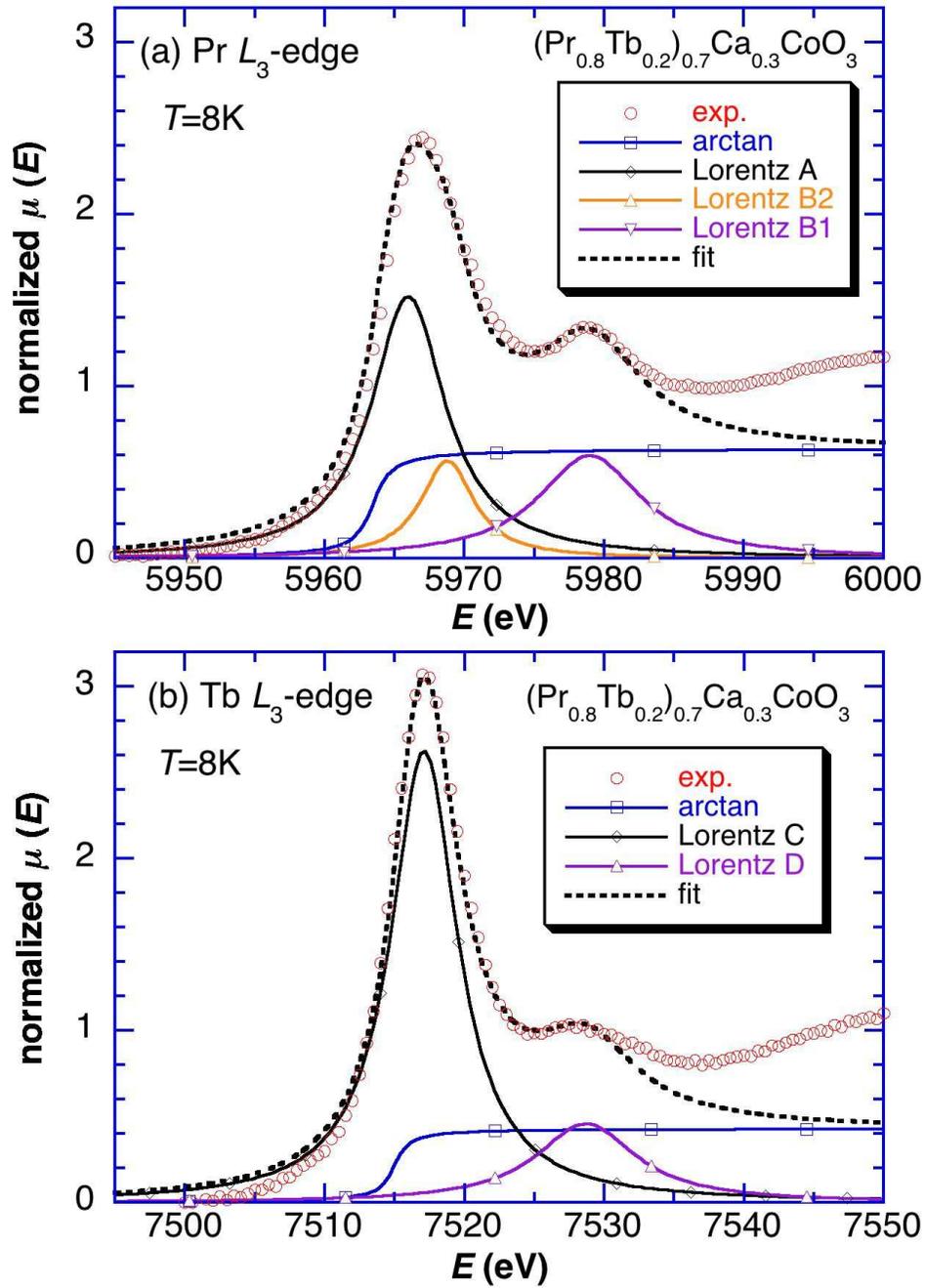}
\caption{Examples of the fitting of the XANES spectrum at the (a) Pr $L_{3}$- and (b) Tb $L_{3}$-edge for the (Pr$_{0.8}$Tb$_{0.2})_{0.7}$Ca$_{0.3}$CoO$_{3}$  sample at 8 K. 
For the fitting of the Pr $L_{3}$-edge, one arctangent function and three Lorentzian functions (A, B1 and B2) are used. For the fitting of the Tb $L_{3}$-edge, one arctangent function and two Lorentzian functions (C and D) are used.
}
\label{fig:4}
\end{figure}

\newpage

\begin{figure}[tb]
\includegraphics*[width=0.8\linewidth]{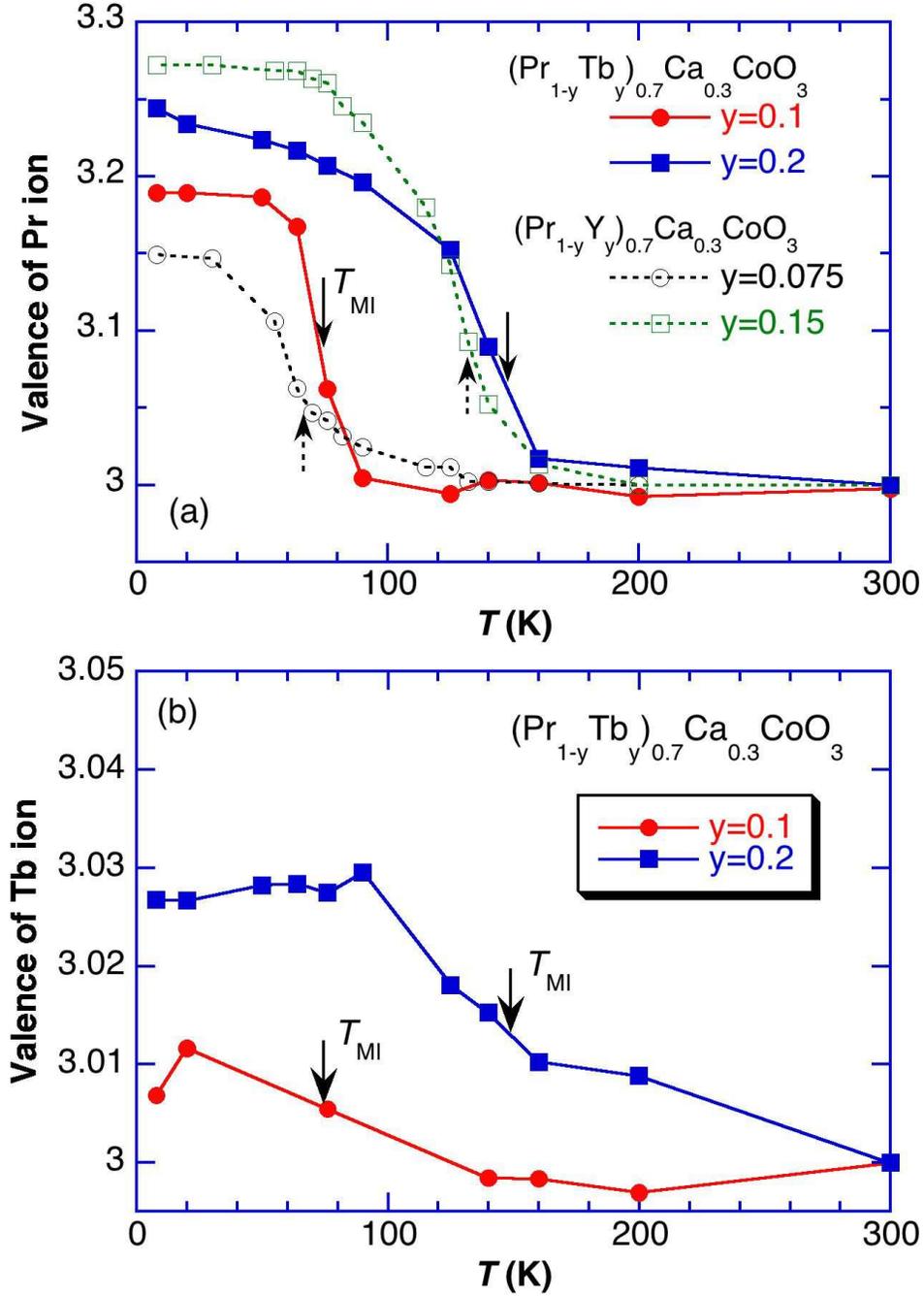}
\caption{Temperature dependence of the average valence of (a) Pr ion and (b) Tb ion for the (Pr$_{1-y}$Tb$_{y})_{0.7}$Ca$_{0.3}$CoO$_{3}$ samples estimated using the XANES spectra and curve fitting. In (a), the estimated valence of Pr ion for the reported (Pr$_{1-y}$Y$_{y})_{0.7}$Ca$_{0.3}$CoO$_{3}$ samples ($y$=0.075 and 0.15)~\cite{Fujishiro2012} is also presented. The uncertainty of the estimated valence values is within 1$\%$ of the absolute value.
}
\label{fig:5}
\end{figure}

\newpage

\begin{figure}[tb]
\includegraphics*[width=0.8\linewidth]{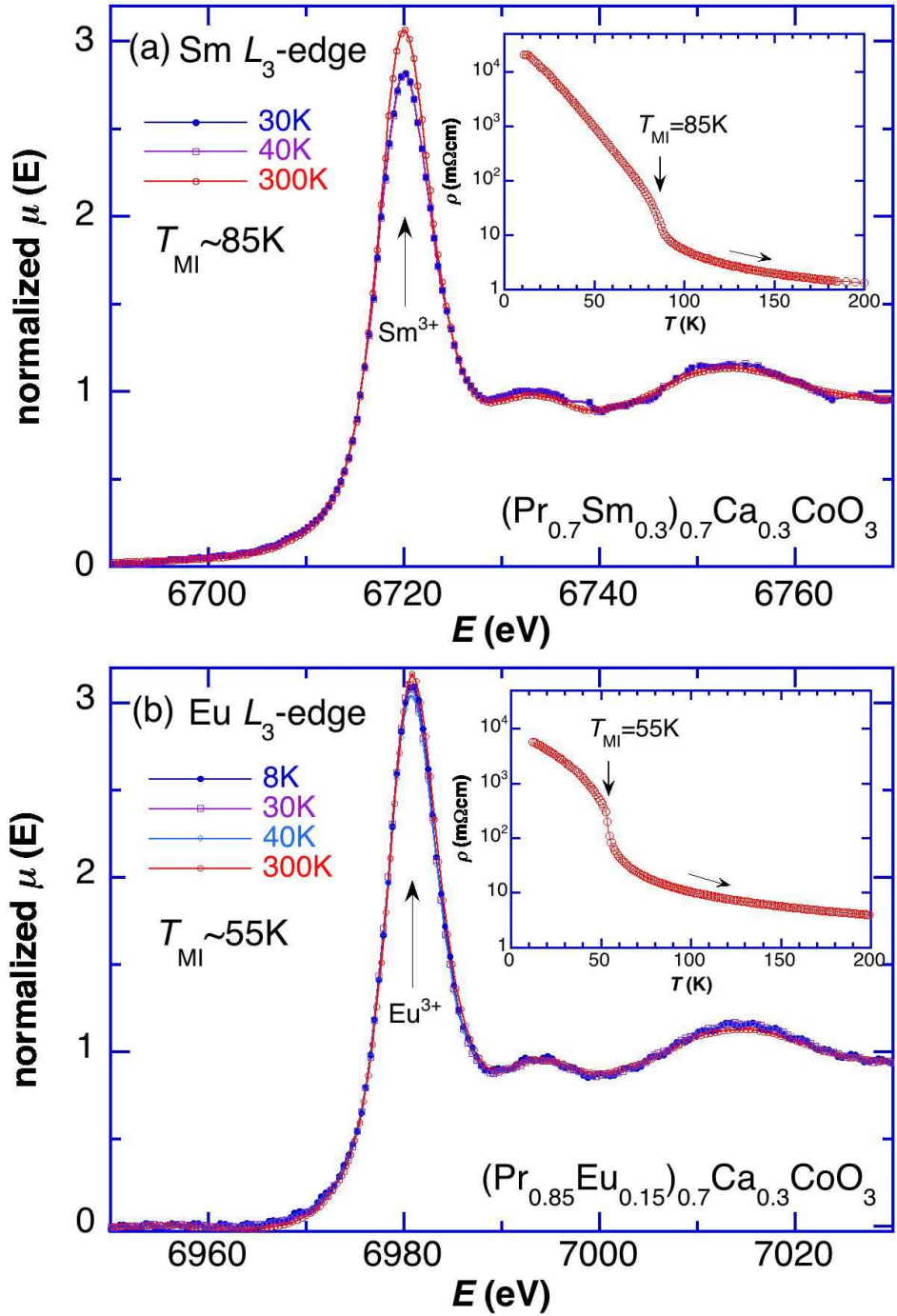}
\caption{Temperature dependence of the XANES spectra at (a) Sm $L_{3}$-edge for (Pr$_{0.7}$Sm$_{0.3})_{0.7}$Ca$_{0.3}$CoO$_{3}$ and (b) Eu $L_{3}$-edge for (Pr$_{0.85}$Eu$_{0.15})_{0.7}$Ca$_{0.3}$CoO$_{3}$ samples.
}
\label{fig:6}
\end{figure}

\newpage

\begin{figure}[tb]
\includegraphics*[width=0.8\linewidth]{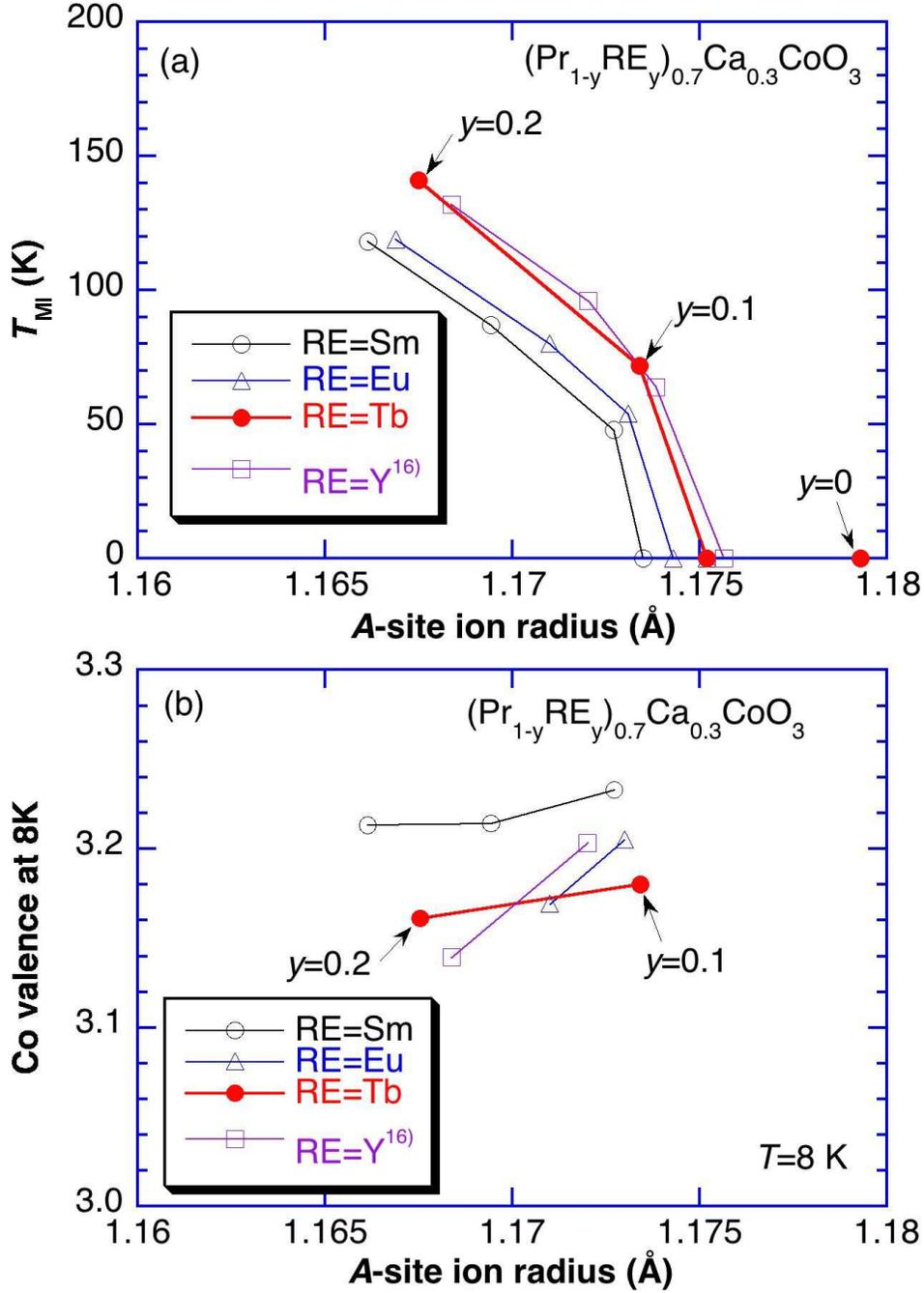}
\caption{(a) The transition temperature $T_{\rm MI}$ and (b) the calculated valence of Co ion at 8 K, as a function of the average ionic radius of the perovskite $A$-site, $\langle r_{A} \rangle$ for various (Pr$_{1-y}$RE$_{y})_{0.7}$Ca$_{0.3}$CoO$_{3}$ systems. The results for the RE=Y were cited from ref.16. In (b), the uncertainty of the estimated valence values is within 1$\%$ of the absolute value.
}
\label{fig:7}
\end{figure}

\end{document}